\documentclass[aps,prl,twocolumn,noshowpacs,superscriptaddress]{revtex4-1}
\usepackage{xspace}
\usepackage{bm,physics}
\usepackage{graphicx}
\usepackage{xcolor,url}
\usepackage{amsmath,amssymb,amsfonts,amsthm}
\usepackage[colorlinks=true,linkcolor=blue,anchorcolor=red,citecolor=blue,urlcolor=blue]{hyperref}
\def \T {\mathcal{T}}
\def \M {\mathcal{M}}
\def \G {\mathsf{G}}
\def \K {\hat{\mathcal{K}}}
\def \I {\hat{I}}
\def \k {\bm{k}}
\def \Z {\mathbb{Z}}

\def \H {\mathcal{H}}

\begin{document}
	
\title{Spinless Mirror Chern Insulator from Projective Symmetry Algebra }

\author{L. B. Shao}
%\email[These authors contributed equally to this work.]{}
\affiliation{National Laboratory of Solid State Microstructures and Department of Physics, Nanjing University, Nanjing 210093, China}
\affiliation{Collaborative Innovation Center of Advanced Microstructures, Nanjing University, Nanjing 210093, China}

\author{Z. Y. Chen}
%\email[These authors contributed equally to this work.]{}
\affiliation{National Laboratory of Solid State Microstructures and Department of Physics, Nanjing University, Nanjing 210093, China}

\author{Kai Wang}
\affiliation{National Laboratory of Solid State Microstructures and Department of Physics, Nanjing University, Nanjing 210093, China}
%\affiliation{Collaborative Innovation Center of Advanced Microstructures, Nanjing University, Nanjing 210093, China}

\author{Shengyuan A. Yang}
\affiliation{Research Laboratory for Quantum Materials, Singapore University of Technology and Design, Singapore 487372, Singapore}

\author{Y. X. Zhao}
\email[]{zhaoyx@nju.edu.cn}
\affiliation{National Laboratory of Solid State Microstructures and Department of Physics, Nanjing University, Nanjing 210093, China}
\affiliation{Collaborative Innovation Center of Advanced Microstructures, Nanjing University, Nanjing 210093, China}

	\begin{abstract}
		It was commonly believed that a mirror Chern insulator (MCI) must require spin-orbital coupling, since time-reversal symmetry for spinless systems contradicts with the mirror Chern number. So MCI cannot be realized in spinless systems which include the large field of topological artificial crystals. Here, we disprove this common belief. The first point to clarify is that the fundamental constraint is not from spin-orbital coupling but the symmetry algebra of time reversal and mirror operations. Then, our theory is based on the conceptual transformation that the symmetry algebras will be \textit{projectively} modified under gauge fields. Particularly, we show that the symmetry algebra of mirror reflection and time-reversal required for MCI can be achieved projectively in spinless systems with lattice $\Z_2$ gauge fields, i.e., by allowing real hopping amplitudes to take $\pm$ signs. Moreover, we propose the basic structure, the twisted $\pi$-flux blocks, to fulfill the projective symmetry algebra, and develop a general approach to construct spinless MCIs based on these building blocks. Two concrete spinless MCI models are presented, which can be readily realized in artificial systems such as acoustic crystals.
	\end{abstract}
	
	\maketitle

{\color{blue}\textit{Introduction.}}
The field of topological matter started with the discovery of the quantum Hall effect or the Chern insulator (CI)~\cite{Klitzing1980,Thouless1982,Haldane1988}. A Chern insulator requires the breaking of time reversal ($\T$) symmetry, which poses difficulty for its realization. For example, a strong magnetic field is needed for the quantum Hall effect~\cite{Klitzing1980}; and for the renowned Haldane model~\cite{Haldane1988}, its delicate flux configuration is not easy to achieve in practical systems~\cite{Jotzu2014}. Later, a significant breakthrough is the discovery of symmetry-protected topological insulators without breaking $\T$. The prominent ones include the $\T$-invariant topological insulator~\cite{Kane2005-1,Kane2005-2} and the mirror Chern insulator (MCI)~\cite{Teo2008,Hsieh2012,Tanaka2012}. The $\T$-invariant topological insulator has found realization in many materials, which led to the boom of the entire field in the past fifteen years~\cite{Hasan2010,Qi2011,Bansil2016,Chiu2016}. Meanwhile, MCI initiated the field of crystalline topological states, which is still actively explored today~\cite{Fu2011,Zhang2019,Vergniory2019,Tang2019}.

%As the first example of topological insulators, the realization of a Chern insulator requires the breaking of time-reversal ($\T$) symmetry. Since crystalline systems naturally have $\T$ invariance, this is unsatisfactory. For the quantum hall effect, strong magnetic fields have to be exerted. For the renowned Haldane model~\cite{Haldane1988}, a flux configuration in each hexagonal plaquette is required, which is not easy to be found in crystalline systems, even for artificial crystals. To remedy the then unfortunate situation, new symmetry-protected topological insulators with $\T$ invariance were proposed. The prominent ones include $\T$-invariant topological insulators and the mirror Chern insulator (MCI)~\cite{Teo2008,Hsieh2012,Tanaka2012}. The discovery of $\T$-invariant topological insulators soon led to a boom in the entire field of topological matter~\cite{Hasan2010,Qi2011,Chiu2016}. Particularly, the MCI initiated the now prosperous field of crystalline topological insulators~\cite{Fu2011,Bradlyn2017,Tang2019,Zhang2019}.

There is a common wisdom regarding MCI: With $\T$ invariance, MCI must require spin-orbit coupling (SOC)~\cite{Teo2008,T-invariance_Bernevig,Ando2015}. Hence, MCI can only be realized in spinful systems, but not spinless systems. In other words, $\T$-invariant spinless MCI does not exist. Because of this, while MCI has been realized in electronic systems of several materials~\cite{Hsieh2012,Tanaka2012,Ando2015}, it was believed to be forbidden for artificial systems, such as acoustic/photonic crystals, 
electric-circuit arrays, 
and mechanical network systems, 
as these systems are intrinsically spinless. Qualitatively  different from the symmetry-protected topological many-body wavefunctions, the artificial crystals can effectively simulate the topological band structures of one Bloch particle by their high tunability. However, despite the rapid growth of topological artificial crystals into a huge and active field in recent years~\cite{Yang2015,Li2018,Ma2019,Lu2014,Yang2019,Ozawa2019,Imhof2018,Yu2020,Prodan2009,Roman2015,Huber2016}, MCI has never been achieved in such systems so far.

%For the topology of the mirror Chern insulators, the mirror ($\M$) symmetry through the $x$-$y$ plane is required to be preserved. Another physically essential ingredient for nontrivial topology is the spin-orbital coupling (SOC)~\cite{Teo2008,Ando2015}. The common wisdom is that without SOC or for spinless systems, it is impossible to realize MCIs. While electronic systems intrinsically have SOC, artificial crystals are spinless systems. Hence, a natural proposal for realizing MCIs in artificial crystals is still understandably absent.  Recently, topological artificial crystals, such as acoustic and photonic crystals~\cite{Yang2015,Li2018,Lu2014,Yang2019}, cold atoms in optical lattices~\cite{Atala2014,Zhang2018,Cooper2019}, periodic mechanical systems~\cite{Prodan2009,Roman2015}, and even electric-circuit arrays~\cite{Imhof2018,Yu2020}, have been developing into a huge and active field. Thus, it is quite unsatisfactory if such an important topological insulator cannot be realized by artificial crystals.

In this Letter, we overturn this common wisdom. We show that spinless MCI does exist, and it can readily realized in artificial systems such as acoustic crystals. This discovery is made possible by advances in two aspects. First, we scrutinize the fundamental requirement for a MCI and clarify that the key factor is not SOC but the symmetry algebra. Explicitly, the only necessary condition is that the mirror operator $\mathcal{M}$ must satisfy $\M^2=-1$ for MCI, provided that it commutes with $\T$. Second, for spatial symmetries such as $\M$, the algebra can be controlled by 
implementing lattice gauge field. Particularly, we show that spinless MCI can be achieved by the simple $\Z_2$ gauge field, meaning that the hopping amplitudes are allowed to take $\pm$ signs, which is something that can be readily engineered in artificial crystals~\cite{Dalibard2011,Ozawa2019,Xue2020,Chunyin_PRL,ni2020demonstration,Xue2021,Li2021}. 
Under gauge fields, symmetries would satisfy so-called projective algebra, which can be designed to meet the requirement of MCI. 
The notion of projective symmetry algebra was applied to physics initially in the study of quantum spin liquids~\cite{Wen2002}. Its profound implications for topological states were only revealed very recently~\cite{Zhao2020,Zhao2021,Shao2021}, and some predictions have already been successfully verified in experiments with acoustic crystals~\cite{Xue2021,Li2021}. Here, we find that
distinct from the previous cases, MCIs need an essentially different mechanism and lattice design. We propose a general prescription to construct MCIs based on a twisted $\pi$-flux block. We show that for any CI model, we can generate an associated spinless MCI using the twisted blocks. We explicitly demonstrate our method via two famous CI models, namely the triangular-lattice model and the Hofstadter model. Our proposed MCI designs can be easily realized in artificial crystals.

{\color{blue}\textit{Symmetry algebras of MCI.}}
Let us start by analyzing the symmetry algebra required for MCI. There are $2$D and $3$D MCIs. For a 2D MCI, the mirror symmetry is with respect to the 2D plane of the system.
For a $3$D MCI, the mirror Chern number is defined on a $2$D sub-system, i.e., some mirror-invariant plane, in the Brillouin zone. Hence, without loss of generality, we focus on $2$D MCIs for clarity.

We find that the fundamental symmetry condition for a MCI is the following algebra~\cite{Algebra_Note}:
\begin{equation}\label{Symm_Alg}
	\quad [\T,\M ]=0,\quad \mathcal{M}^2=-1.
\end{equation}
To understand this, we first note that with the $\M$ symmetry, states of the system can be separated into the mirror-even and mirror-odd subspaces. 
Particularly, the momentum-space Hamiltonian $\H(\k)$ can be put into the block diagonal form
\begin{equation}\label{block_H}
	\H(\k)=\begin{bmatrix}
		h_{+}(\k) & 0\\
		0 & h_{-}(\k)
	\end{bmatrix},
\end{equation}
in accord with the two eigenspaces of $\M$. If the algebra \eqref{Symm_Alg} holds, $\M$ has eigenvalues of $\pm i$. 
For eigenstates $|\psi_{\pm}\rangle$ with
$\M |\psi_{\pm}\rangle=\pm i |\psi_{\pm}\rangle$, we observe that $\M\T|\psi_{\pm}\rangle=\T \M|\psi_{\pm}\rangle=\T(\pm i |\psi_{\pm}\rangle)=\mp i\T|\psi_{\pm}\rangle$, since $\T$ is an anti-unitary operator involving complex conjugation. This just means that $\T$ {exchanges} the two eigenspaces. Hence, we must have $uh_{+}^*(-\k)u^\dagger=h_{-}(\k)$ for some unitary matrix $u$ determined by $\T$, i.e., $\T$ transforms $h_{\pm}(\k)$ into $h_{\mp}(\k)$.
Assuming that $\H(\k)$ is gapped, we can calculate Chern numbers $C_{\pm}$ for $h_{\pm}(\k)$, respectively. Since $\T$ inverses the Chern number, $h_{+}(\k)$ and $h_{-}(\k)$ must have opposite Chern numbers: $C_{+}=-C_{-}$. Thus, although the total Chern number $C=C_++C_-=0$, each block $h_{\pm}$ can have a nontrivial Chern number, and accordingly $C_+$ is defined as the mirror Chern number~\cite{Teo2008}.

From the above reasoning, we see that the essence for a nontrivial mirror Chern number is: $\T$ must \emph{exchange} the eigenspaces of $\M$. For electronic systems with SOC, $\T=i\sigma_2\K\I$ and $\M=i\sigma_3$ in the spin space of each electron, where $\sigma$'s are the Pauli matrices. Hence, the algebra in \eqref{Symm_Alg} is naturally satisfied. In comparison, if $\M^2=+1$ and $[\T,\M]=0$, as for typical spinless systems, then $\T$ would preserve the eigenspaces of $\M$. This is because the eigenvalues $\pm 1$ of $\M$ are real numbers that commutes with $\T$. Although we still can write $\H(\k)$ into the block diagonal form \eqref{block_H} for eigenspaces of $\pm 1$, both $h_{\pm}(\k)$ are invariant under $\T$, and therefore they each must have a zero Chern number, i.e., $C_\pm=0$.

We have some remarks before proceeding. First, contrary to common perceptions, the relation $\T^2=\pm 1$ is not essential for MCI, since $\T$ inverses the Chern number in both cases. Second, it is clear from the analysis that the key factor here is not SOC but the symmetry algebra, particularly, whether $\M^2$ equals $+1$ or $-1$. %As we shall see, $\M^2=-1$ can also be achieved in spinless systems. 

\begin{figure}
	\centering
	\includegraphics[scale=0.43]{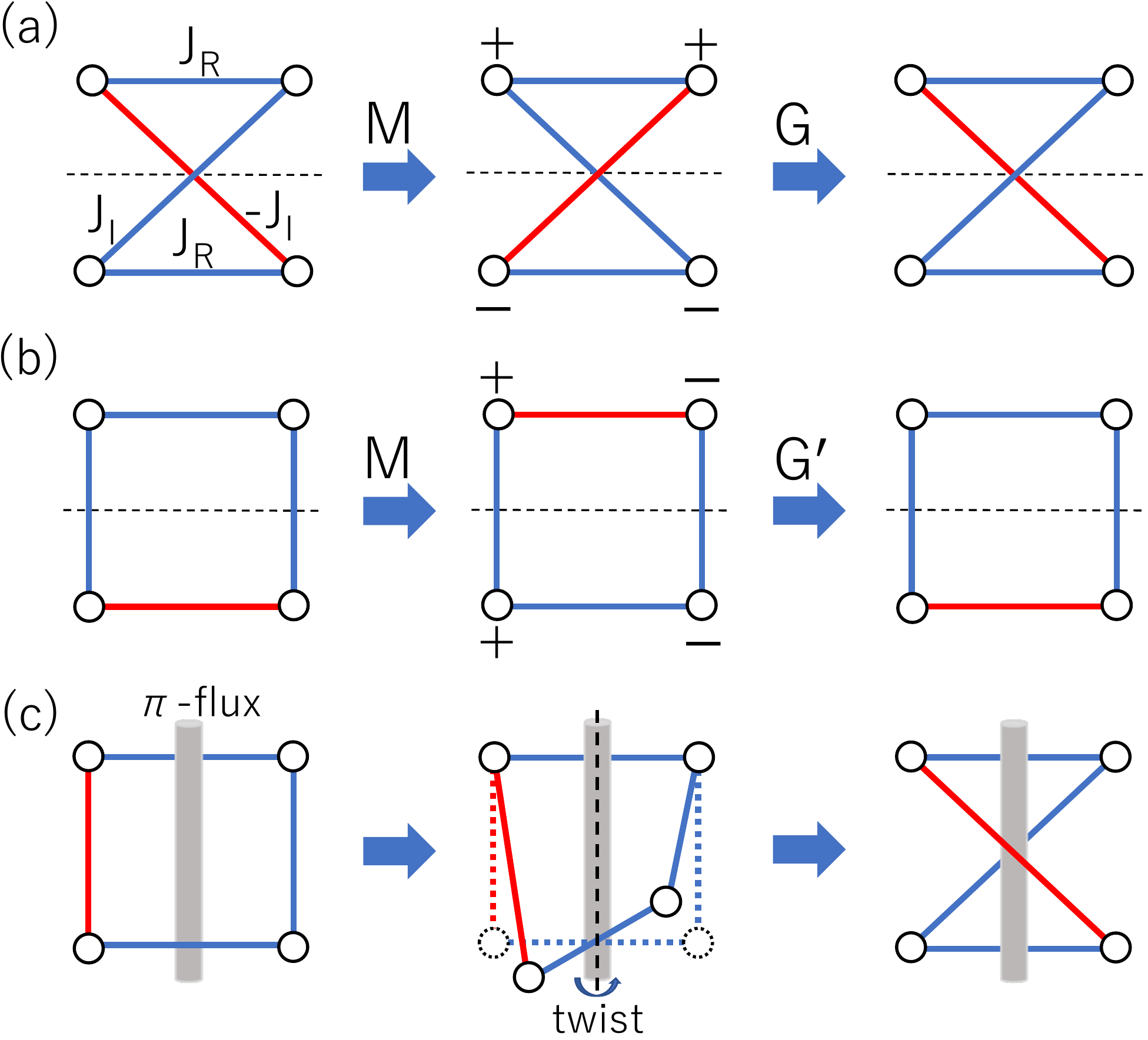}
	\caption{Two four-site tight-binding models. Red and blue lines denote negative and positive hopping amplitudes, respectively. The dashed horizontal lines in (a) and (b) are the reference lines for the spatial mirror refection $M$. The signs on the middle figures specify the gauge transformations $\G$ and $\G'$ for (a) and (b), respectively. The tight-binding models are invariant under $M$ followed by $\G$ or $\G'$. (c) The model in (a) is a twist of a rectangle with flux $\pi$. The $\pi$ flux is depicted by a thin gray tube. \label{Small_Models}}
\end{figure}

{\color{blue}\textit{Projective symmetry algebra.}} Although it seems from experience that spinless systems always have $\M^2=+1$, we show that with certain $\Z_2$ gauge fields the condition in \eqref{Symm_Alg} can be realized as projective symmetry algebra in spinless systems.

Let us start with some general considerations on a mirror symmetric system with a given $\Z_2$ gauge configuration. The field is described by a chosen configuration of gauge connections, i.e., signs $\pm 1$ of real hopping amplitudes. The gauge-connection configuration in general is not invariant under the spatial mirror reflection $M$, but will be changed to another equivalent configuration (another gauge choice), which is related to the original one by a $\Z_2$ gauge transformation $\G$. $\G$ is specified by assigning a sign of $+1$ or $-1$ to the basis at each site. Then, the physical mirror operator will be represented as the combination
\begin{equation}\label{Phys_M}
	\M=\G M,
\end{equation}
namely the spatial reflection $M$ followed by the gauge transformation $\G$. Since both $M$ and $\G$ in real space are real matrices, $[\M,\T]=0$ in \eqref{Symm_Alg} is trivially satisfied. Moreover, since $M^2=\G^2=1$, to satisfy $\M^2=-1$ in \eqref{Symm_Alg}, we need the anti-commutation relation between $\G$ and $M$,
\begin{equation} \label{Anti-comm}
	\{\G,M\}=0.
\end{equation}
The anti-commutativity is equivalent to $M\G M^{-1}=-\G$, which just means that $M$ inverses all signs at all sites for $\G$. This observation is a guiding principle for the model construction.

Let us consider a simple system consisting of only four sites, as illustrated in Fig.~\ref{Small_Models}. Positive and negative hopping amplitudes are marked with blue and red colors, respectively. For the model in Fig.~\ref{Small_Models}(a), the Hamiltonian is given by
\begin{equation}\label{twist_H}
	\H=J_R \tau_0\otimes \sigma_1 + J_I\tau_2\otimes\sigma_2,
\end{equation}
where $\tau$'s and $\sigma$'s be two sets of Pauli matrices which respectively operate on the row and column indices of the block. The mirror reflection $M=\tau_1\otimes\sigma_0$ through the dashed horizontal line exchanges diagonal and anti-diagonal hopping processes, and therefore the gauge connections are changed. To restore the original gauge connections, the gauge transformation $\G=\tau_3\otimes\sigma_0$ needed is specified in the middle figure of Fig.~\ref{Small_Models}(a). Obviously, $M$ inverses $\G$ with $\{M,\G\}=0$. Hence, Eq.~\eqref{Anti-comm} is satisfied, and the resulting projective symmetry algebra will be \eqref{Symm_Alg} needed for the MCI, where
\begin{equation}\label{Mirror}
	\M=\G M=i\tau_2\otimes \sigma_0.
\end{equation}
On the other hand,  the setup, $\H^{(b)}=J_1 \tau_1\otimes\sigma_0+J_2 \tau_3\otimes\sigma_1$, in Fig.~\ref{Small_Models}(b) does not work. In the middle figure of Fig.~\ref{Small_Models}(b), the gauge transformation $\G'=\tau_0\otimes\sigma_3$ is invariant under $M$, and therefore $[\G',M]=0$ rather than $\{\G',M\}=0$, and we have $\M'^2=+1$ with $\M'=\tau_1\otimes\sigma_3$ in this case.

\begin{figure}
	\centering
	\includegraphics[scale=0.59]{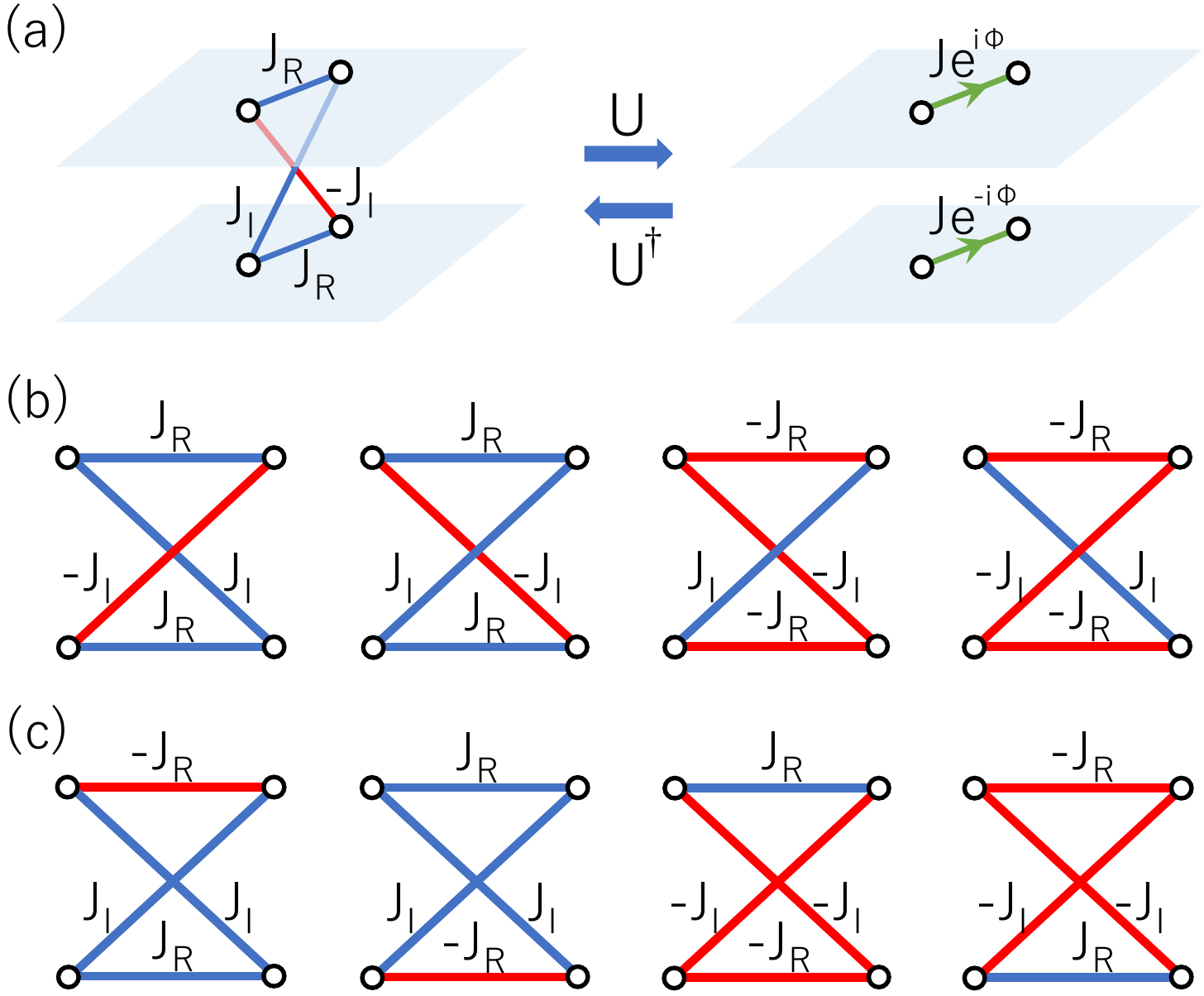}
	\caption{(a) The equivalence mapping between the twisted $\pi$-flux block and two complex hopping amplitudes. $U$ is the unitary transformation diagonalizing $\M$. %$U$ can transform the twisted $\pi$-flux block consisting of only real hopping amplitudes to two complex hopping amplitudes $Je^{\pm i\phi}$. Here, $Je^{i\phi}=J_R+iJ_I$ with $J>0$. $U^\dagger$ does the inverse transformation. 
		(b) Four twisted $\pi$-flux blocks with $\M=i\tau_2\otimes\sigma_0$. (c) The other four with $\M=i\tau_2\otimes\sigma_3$.
		\label{Mapping}}
\end{figure}

The arrangement  in Fig.~\ref{Small_Models}(a) is referred to as a twisted $\pi$-flux block, because, as shown in Fig.\ref{Small_Models}(c), it is a twist of a $\pi$-flux rectangle with $\H^{(c)}=-J_1\tau_1\otimes\sigma_3+J_2\tau_0\otimes\sigma_1$. It is clear that both the $\pi$-flux and the twist are essential for achieving the symmetry algebra \eqref{Symm_Alg}. 
In fact, there are eight possible $\Z_2$ gauge-connection configurations on this twisted $\pi$-flux block [see Fig.~\ref{Mapping}(b) and (c)], which are equivalent to each other through some $\Z_2$ gauge transformations, and therefore all can realize the projective algebra \eqref{Symm_Alg}. %We call them twisted $\pi$-flux blocks.

%to describe the $\pi$-flux [see Fig.\ref{Mapping}(b) and (c)], which are equivalent to each other through $\Z_2$ gauge transformations. Importantly, the projective symmetry algebra \eqref{Symm_Alg} is invariant under $\Z_2$ gauge transformations. Hence, all the eight twisted $\pi$-flux blocks shown in Fig.\ref{Mapping}(b) and (c) can realize the projective symmetry algebra.

{\color{blue}\textit{General method for constructing spinless MCI.}} We develop a general method for constructing spinless MCIs. Our construction is based on the twisted $\pi$-flux block.
Since the blocks in Fig.~\ref{Mapping}(b) and \ref{Mapping}(c) are gauge equivalent, it is sufficient to look into one of them. Consider the one in Fig.~\ref{Small_Models}(a) with the Hamiltonian \eqref{twist_H} and mirror operator \eqref{Mirror}.

%Then, we read off from Fig.~\ref{Small_Models}(a) that $\G=\tau_3\otimes\sigma_0$ and $M=\tau_1\otimes\sigma_0$. From Eq.~\eqref{Phys_M}, we find
%\begin{equation}\label{Mirror}
%	\M=i\tau_2\otimes \sigma_0.
%\end{equation}
%Clearly, $\T=\K$ for spinless systems. Now, $\M$ and $\T$ of the spinless system projectively follow the the desired symmetry algebra \eqref{Symm_Alg} which was thought to apply only for spinful systems. The Hamiltonian for the block  is given by
%\begin{equation}
%	\H=J_R \tau_0\otimes \sigma_1 + J_I\tau_2\otimes\sigma_2,
%\end{equation}
%which clearly commutes with $\T$ and $\M$. It is worth noting that vertical hopping terms, $\tau_1\otimes\sigma_{0,3}$, are excluded from the model by the mirror symmetry $\M$.

It is enlightening to view the model in the eigenspaces of $\M$, which can be achieved by performing the unitary transformation $U=\exp(-i\tau_1\otimes\sigma_0\pi/4)$. Under the transformation, $\M\rightarrow U\M U^\dagger=i\tau_3\otimes\sigma_0$.
Now, $\tau_3$ corresponds to the index of the two eigenspaces of $\M$ with eigenvalues $\pm i$, and the Hamiltonian is transformed into the block diagonal form,
\begin{equation}\label{D_Small_Model}
	U\H U^\dagger=\begin{bmatrix}
		J_R\sigma_1+J_I\sigma_2 & 0\\
		0 & J_R\sigma_1-J_I\sigma_2
	\end{bmatrix}_\tau.
\end{equation}
Meanwhile, $\T$ is transformed to be $U\T U^\dagger=-i\tau_1\K$, which, as expected, exchanges the two diagonal blocks of the Hamiltonian, i.e., the two eigenspaces of $\M$. We refer to the two diagonal blocks as eigenvalue layers. Then, $J_R\sigma_1\pm J_I\sigma_2$ can be interpreted as hopping amplitudes $Je^{\pm i\phi}$ between two lattice sites in each  eigenvalue layer, as illustrated in Fig.~\ref{Mapping}(a). The hopping phase and strength are explicitly given by
\begin{equation}\label{phase}
	e^{i\phi}=(J_R+i J_I)/J,\quad J=\sqrt{J^2_R+J_I^2}.
\end{equation}

%In Fig.\ref{Mapping}, the red-blue twisted block is equivalent to $Je^{\pm i\phi}$ on the upper and lower layers, respectively. If we want $Je^{\mp i\phi}$, respectively, for the upper and lower layers, we just need to use the blue-red twisted block, i.e., to exchange diagonal and anti-diagonal hopping amplitudes.

It is important that starting from a spinless model with purely real hopping amplitudes, we are able to convert it to a system with complex hopping amplitudes. Conversely, for any prescribed complex hopping amplitude, we can construct a spinless twisted $\pi$-flux block such that one of its eigenvalue layer realizes the amplitude.
Note that the hopping phase is a key ingredient for CI models. For instance, the Haldane model is characterized by the second neighbor hopping phase $\phi$ on a honeycomb lattice~\cite{Haldane1988,shen2012topological,Bernevig2013}. For the Hofstadter model, each square plaquette has a flux $\phi$~\cite{Hofstadter1976,Kohmoto1985}. 
%And in the triangular-lattice CI model, the nearest-neighbor hopping amplitudes along one direction have a phase $\phi$.

Based on this understanding, we have the following general method to construct spinless MCIs from any CI model. 
Given such a CI model $H_C(\phi)$ with a characteristic phase $e^{i\phi}$, we can construct a spinless MCI $H_{MC}(\phi)$ protected by $\M$ and $\T$ by the invertible mapping illustrated in Fig.~\ref{Mapping}(a). Specifically, we take $H_C(\phi)$ and $H_C(-\phi)$ as two independent layers. Then, we perform the mapping in Fig.~\ref{Mapping}(a) inversely, i.e.,  replace each pair of complex hopping amplitudes with phases $\pm\phi$ on the two layers by a twisted $\pi$-flux block. The resultant bilayer tight-binding model is just the wanted spinless MCI. For this system, the mirror operator is $\M=i\tau_2$ with $\tau$'s operating on the layer degrees of freedom, and $\T=\K$.

The corresponding mapping for the momentum-space Hamiltonians can be immediately constructed. Let $h_C(\k,\phi)$ be the momentum-space Hamiltonian for $H_C(\phi)$. Then, we introduce
\begin{equation}
	h_{\pm}(\k,\phi)=\frac{1}{2}[h_C(\k,\phi)\pm h_C(\k,-\phi)].
\end{equation}
$h_{+}(\k,\phi)$ [$h_{-}(\k,\phi)$] is an even (odd) function of $\phi$. Since $h_C(\k,\phi)$ and $h_C(\k,-\phi)$ are related by $\T$ symmetry, $h_{+}(\k,\phi)$ [$h_{-}(\k,\phi)$] are also even (odd) under $\T$ operation. Then, the $\T$-invariant Hamiltonian $\H_{MC}(\k,\phi)$ for the bilayer MCI is given by
\begin{equation}\label{H_MC}
	\H_{MC}(\k)=\begin{bmatrix}
		h_+(\k,\phi) &  -i h_{-}(\k,\phi)\\
		i h_{-}(\k,\phi) & h_{+}(\k,\phi)
	\end{bmatrix}_{\tau}.
\end{equation}
Finally, we should substitute $\phi$ by $J_R$ and $J_I$ according to Eq.~\eqref{phase}. Here, $\M=i\tau_2$, and $\T=\K\I$ with $\I$ the inversion of momenta.

\begin{figure}
	\centering
	\includegraphics[scale=0.33]{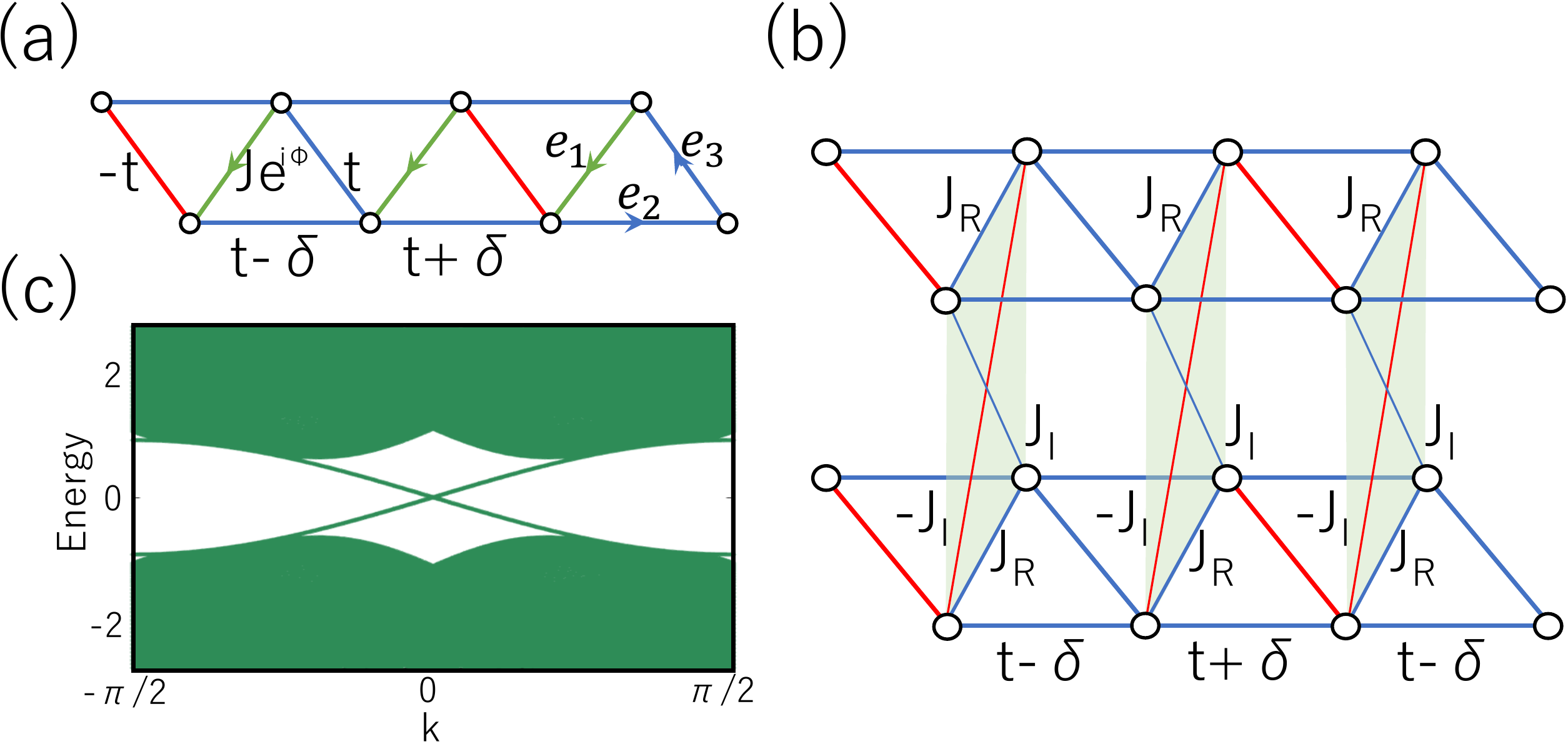}
	\caption{(a) The triangular-lattice CI model. Green, red and blue bonds denote complex, negative and positive hopping amplitudes, respectively. The model is dimerized by $\delta$ along $\bm{e}_2$. (b) The corresponding MCI model.  Complex hopping amplitudes in (a) are replaced by twisted $\pi$-flux blocks.  (c) The spectrum of the bilayer model with edges along $\bm{e}_2$. The parameter values are chosen as $t=1,\delta=0.5,J_R=J\cos \phi,J_I=J\sin\phi$ with $J=2$ and $\phi=2\pi/5$~\cite{Note_Flux}. \label{Triangle}}
\end{figure}

In the above discussion, we have chosen the particular twisted $\pi$-flux block in Fig.~\ref{Small_Models}(a) to demonstrate our idea. Other blocks in Fig.~\ref{Mapping}(b) and \ref{Mapping}(c) can also be used. It is not difficult to see that these eight blocks can be put into two groups, with members in a group sharing the same representation of $\M$: For those in Fig.~\ref{Mapping}(b), $\M=i\tau_2\otimes \sigma_0$; whereas for Fig.~\ref{Mapping}(c), $\M=i\tau_2\otimes \sigma_3$~\cite{SM}. It should be noted that if one uses more than one kind of blocks in a single model, then these blocks should be chosen from only one of the two groups to maintain the same representation for $\M$. 

Nevertheless, we emphasize that generically chiral edge states of the two mirror layers in Fig.\ref{Mapping}(a) will be simultaneously excited in real space, since mirror reflection relates two real-space layers.

%find that all the four blocks in Fig.\ref{Mapping}(b) have the same $\M$, and so do the other four in Fig.\ref{Mapping}(c). Hence, we can freely choose multiple blocks from the four in Fig.\ref{Mapping}(b) or in (c) to connect two layers. But, we are not allowed to choose from both groups.
%
%They can be further put into two groups, with members in a group sharing the same representation of $\M$, as shown in Fig.~\ref{Mapping}(b) and \ref{Mapping}(c).

{\color{blue}\textit{Concrete models}.} We demonstrate our theory by constructing two concrete spinless MCIs from well-know CI models.

The first one is based on the triangular-lattice CI model illustrated in Fig.~\ref{Triangle}(a). Let us denote the three bond vectors as $\bm{e}_a$ with $a=1,2,3$. The hopping amplitudes along $\bm{e}_1$ are $Je^{i\phi}$, and the phase $e^{i\phi}$ characterizes the model. The other hopping amplitudes are real, and those along the $\bm{e}_3$ bonds change signs alternatively in the $\bm{e}_2$ direction. The hopping amplitudes along $\bm{e}_3$ is $t$, and the hoppings along $\bm{e}_2$ have a dimerized pattern with amplitudes $t\pm \delta$. If $J>t$ and $\delta\ne 0$, the model can realize a CI by tuning $\phi$, and the Chern number $C=\pm 1$~\cite{SM}. The corresponding spinless MCI model is illustrated in  Fig.~\ref{Triangle}(b). Following our general method, it is a bilayer model with each original bond with phase $\phi$ replaced by a twisted $\pi$-flux block, and the Hamiltonian is in the form of \eqref{H_MC}~\cite{SM}.
%Particularly, $J_R$ and $J_I$ for the twisted blocks connecting two layers are chosen according to \eqref{phase}. According to \eqref{H_MC}, the Hamiltonian $\H^{\mathrm{tr}}(\k)$ is given by
%\begin{equation}
%\H^{\mathrm{tr}}(\k)=\tau_0\otimes h_{+}^{\mathrm{tr}}(\k)+\tau_2\otimes h_{-}^{\mathrm{tr}}(\k)
%\end{equation}
%where
%\begin{eqnarray}
%		h_+^{\mathrm{tr}}(\k)&=&2(J_R\cos\k\cdot\bm{e}_1+t\cos\k\cdot\bm{e}_2)\sigma_1 \nonumber\\
%		&&+2\delta\sin\k\cdot\bm{e}_2\sigma_2-2t\cos\k\cdot\bm{e}_3\sigma_3,\\
%		h_-^{\mathrm{tr}}(\k)&=&2J_I\sin\k\cdot\bm{e}_1\sigma_1.\nonumber
%\end{eqnarray}
%Here, $\tau$'s act on the layer index, and $\sigma$'s operates in the sublattice space of each layer. As expected, the mirror operator $\M=i\tau_2\otimes\sigma_0 $ commutes with $\H(\k)$. We choose values of $J_{R}$ and $J_I$ in accord with $Je^{i\phi}$, and the other parameters take the same values as those of the triangular model. With the parameter values of Fig.\ref{Triangle} (c) and (d), the mirror Chern number $C_{+}=C=1$.
The spectrum with the open boundary conditions for an edge along $\bm{e}_2$ is shown in Fig.~\ref{Triangle}(c). We see a pair of left-handed and right-handed topological edge bands, which correspond to the unit mirror Chern number.

%simultaneously diagonalizing $\M$ and $\H(\k)$, we find that each diagonal block of $\H(\k)$ has nontrivial Chern numbers $C_+=-C_-=1$.

\begin{figure}
	\centering
	\includegraphics[scale=0.29]{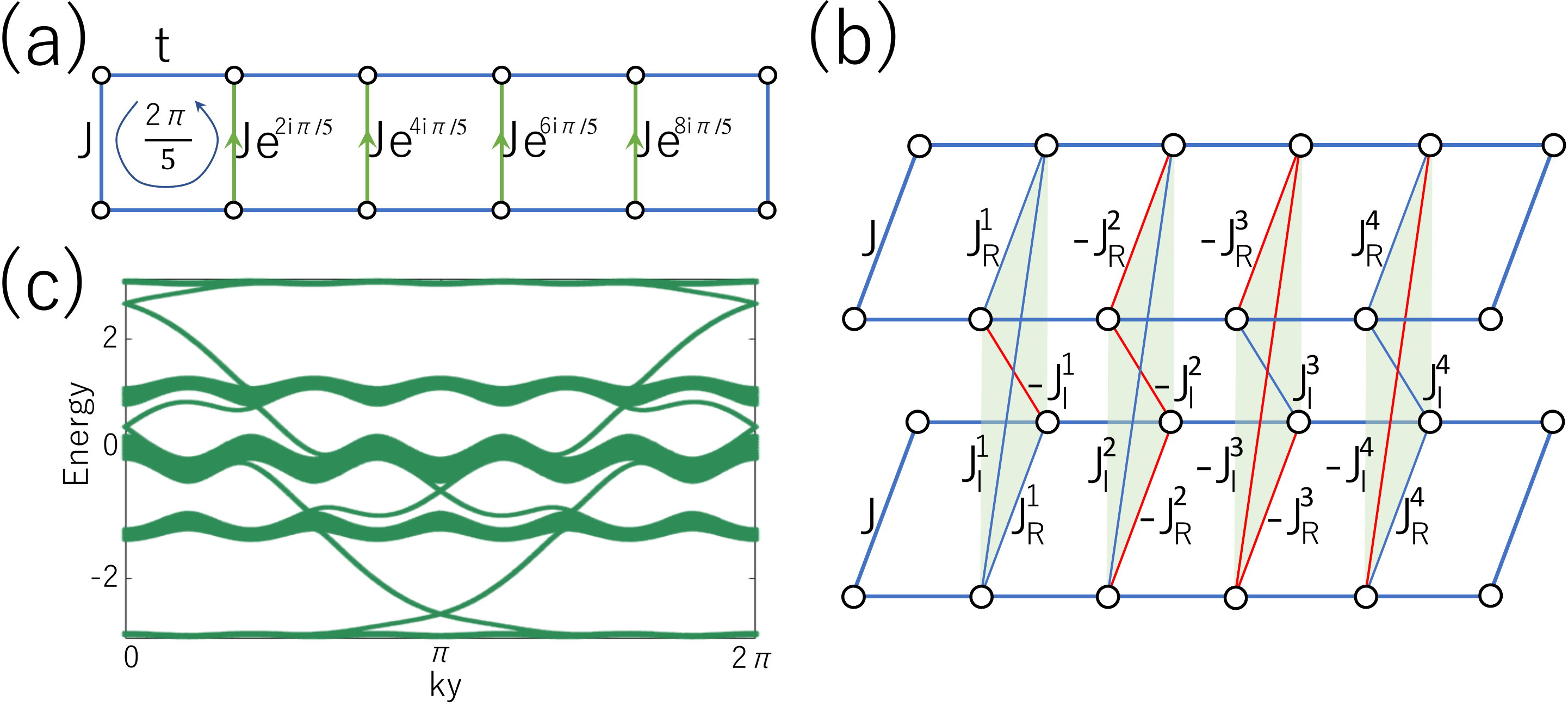}
	\caption{(a) The Hofstadter model. The flux $\phi$ is $2\pi/5$. Green, red and blue bonds denote complex, negative and positive hopping amplitudes, respectively. (b) The corresponding Hofstadter MCI model. The green bonds are replaced by the four twisted $\pi$-flux blocks in Fig.\ref{Mapping}(b). (c) The spectrum of the MCI model with edges along the $y$ direction. The parameter values are chosen as $t=J=1$, and accordingly $ J_R^n=J|\cos 2n\pi/5| $ and $ J_I^{n}=J|\sin2n\pi/5| $ with $ n=1,2,3,4 $~\cite{Note_Flux}. \label{Square}}
\end{figure}

The second example is based on the renowned Hofstadter CI model~\cite{Hofstadter1976,Kohmoto1985}. It is just a square-lattice model with flux $\phi$ per square plaquette. Here, we choose $\phi=2\pi/5$. The unit cells for the Hofstadter model and the corresponding MCI model are illustrated in Fig.~\ref{Square}(a) and (b). We choose to connect the two layers of the MCI by using the four different twisted $\pi$-flux blocks in Fig.~\ref{Mapping}(b), so that the required flux in each eigenspace layer of $\M$ can be realized~\cite{SM}. The model has four energy gaps separating energy bands into five groups. Ordered by energy, the mirror Chern numbers for these five groups of bands are found as $-1$, $-1$, $4$, $-1$ and $-1$~\cite{SM}. The topological chiral edge bands emerge accordingly, as shown in Fig.~\ref{Small_Models}(c). Particularly, for the middle two gaps, each hosts two left-handed and two right-handed chiral bands, since the sums of mirror Chern numbers below them are $\pm 2$, respectively.

{\color{blue}\textit{Discussion.}} We have disproved the common belief that MCI must require SOC and hence can only be realized in spinful systems.
%Although only two models are constructed here, it is clear that the method is general. With the demonstrated method, it is straightforward to transform other CI models, e.g., the Haldane model, to a spinless MCI preserving $\T$ and $\M$.
Essentially, we have clarified the most fundamental symmetry requirement for MCIs and how $\Z_2$ gauge fields can projectively modify the symmetry algebra to achieve the requirement in spinless systems. 
Although we have focused on $2$D spinless MCIs, the discussion can be directly extended to $3$D spinless MCI models.

Our work greatly broadens the experimental relevance of MCIs. Our proposed spinless MCI models can be readily realized in acoustic crystals with engineerable $\Z_2$ gauge fields~\cite{Xue2020,Chunyin_PRL,ni2020demonstration,Xue2021,Li2021}. Particularly, an acoustic realization of the twisted $\pi$-flux blocks is given in detail in the SM~\cite{SM}.
Other artificial crystals, such as cold atoms in optical lattices~\cite{Zhang2018,Cooper2019} and electric-circuit arrays~\cite{Imhof2018,Yu2020}, may also be possible platforms for realizing our proposals.

%{\color{blue}\textit{Summary}}

\bibliographystyle{apsrev4-1}
\bibliography{Mirror_Ref}

\begin{thebibliography}{50}
\bibitem [{Alg()}]{Algebra_Note}%
  \BibitemOpen
  \href@noop {} {}\bibinfo {note} {One may redefine the mirror operator as
  $\tilde{\M}=-i\M$. Since $\T$ is anti-unitary, the algebra in
  \eqref{Symm_Alg} would be translated to an equivalent form of
  $\{\T,\tilde{\M}\}=0$ and $\tilde{\M}^2=1$. In the following discussion, we
  shall stick to the form in \eqref{Symm_Alg}.}\BibitemShut {Stop}%
\bibitem [{Not()}]{Note_Flux}%
  \BibitemOpen
  \href@noop {} {}\bibinfo {note} {The parameters are not required to be
  accurate. The topological phase is robust under the deformation of parameters
  as long as the mirror symmetry is preserved and the energy gap is not
  closed.}\BibitemShut {Stop}%
\end{thebibliography}

\newpage

\onecolumngrid
\renewcommand{\theequation}{S\arabic{equation}}
\setcounter{equation}{0}
\renewcommand{\thefigure}{S\arabic{figure}}
\setcounter{figure}{0}

\def \Z {\mathbb{Z}}
\def \T {\mathcal{T}}
\def \H {\mathcal{H}}
\def \e {\bm{e}}
\def \k {\bm{k}}
	
\section{Supplemental Material for \\ ``Spinless Mirror Chern Insulator from Projective Symmetry Algebra"}

\section{Two groups of Twisted $ \pi $-Flux Blocks}
	
\begin{figure}[t]
	\centering
	\includegraphics[scale=1]{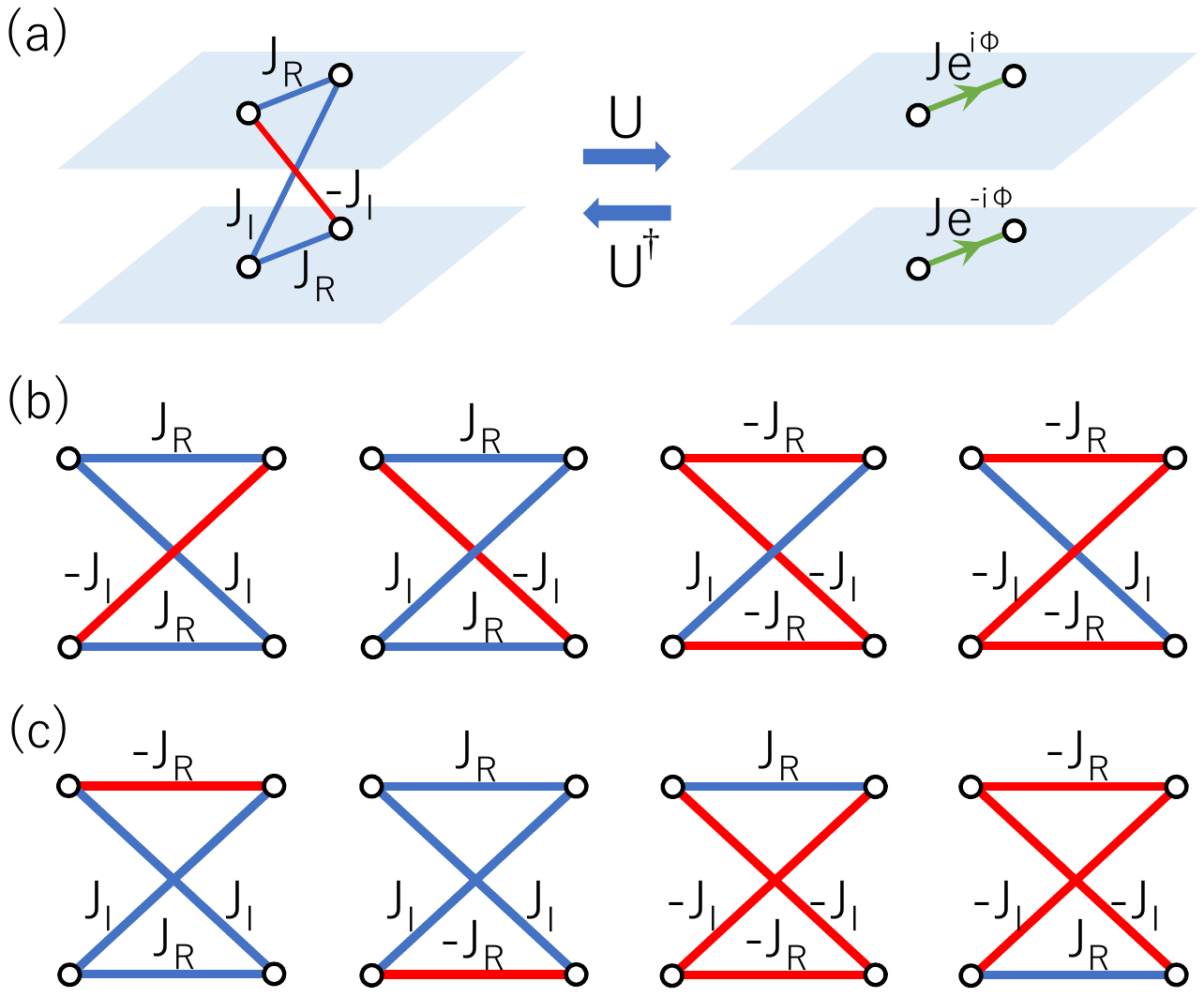}
	\caption{(a) The mapping from the twisted $ \pi $-flux block to the block diagonal form. (b) and (c) contain the twisted $ \pi $-flux blocks symmetric under the mirror symmetry $ \mathcal{M}_1=\mathsf{G}_1M $ and $ \mathcal{M}_2=\mathsf{G}_2M $, respectively. ~\label{SM-Mapping}}
\end{figure}

The eight twisted $\pi$-flux blocks are essentially important for our general method for model construction. In this section, we derive basic results and supply technical details for the eight twisted $\pi$-flux blocks. They are divided into two four-block groups, each with a projective mirror operator. The following materials in this section is organized in accord with the two-group division. For convenience, we quote Fig.~2 of the main text here as Fig.~\ref{SM-Mapping}.

The four twisted $ \pi $-flux blocks in Fig.~\ref{SM-Mapping}(b) is invariant under $	\mathcal{M}_1=\mathsf{G}_1M $, where
\begin{equation}%\label{key}
	\mathsf{G}_1=\tau_3\otimes\sigma_0,\qquad M=\tau_1\otimes\sigma_0.
\end{equation}
	Obviously, they satisfy the anti-commutation relation $ \{\mathsf{G}_1,M\}=0 $, and
	\begin{equation}
		\mathcal{M}_1=i\tau_2\otimes\sigma_0.
	\end{equation}
	The operation of the mirror symmetry $ \mathcal{M}_1 $ in this case is illustrated by Fig.~1(a) in the main text. By the unitary transformation $ U=e^{-i\pi\tau_1\otimes\sigma_0/4} $ as given in the main text, $ \mathcal{M}_1 $ can be diagonalized as $ U\mathcal{M}_1U^\dagger= i\tau_3\otimes\sigma_0 $. Fig.~\ref{SM-Mapping}(a) is just the second block of Fig.~\ref{SM-Mapping}(b), and its Hamiltonian is written as
	\begin{equation}
		\H=J_R\tau_0\otimes\sigma_1+J_I\tau_2\otimes\sigma_2.
	\end{equation}
	By the unitary transformation $ U $, the Hamiltonian can be transformed as
	\begin{equation}%\label{key}
		U\H U^\dagger=\begin{bmatrix}
			J_R\sigma_1+J_I\sigma_2 & \   \\   \ & J_R\sigma_1-J_I\sigma_2
		\end{bmatrix}.
	\end{equation}
Hence, the Hamiltonian $ U\H U^\dagger $ contains two blocks, which can be represented as the right panel of Fig.~\ref{SM-Mapping}(a). We give the detail mapping of the four Hamiltonians for Fig.~\ref{SM-Mapping}(b) in the following:
\begin{equation}%\label{key}
		\begin{split}
			\H&=J_R\tau_0\otimes\sigma_1-J_I\tau_2\otimes\sigma_2\ \ \  \Rightarrow\   U\H U^\dagger=\left( J_R\sigma_1-J_I\sigma_2 \right)\oplus\left(J_R\sigma_1+J_I\sigma_2\right),\\	\H&=J_R\tau_0\otimes\sigma_1+J_I\tau_2\otimes\sigma_2\ \ \  \Rightarrow\  U\H U^\dagger=\left( J_R\sigma_1+J_I\sigma_2 \right)\oplus\left(J_R\sigma_1-J_I\sigma_2\right),\\
			\H&=-J_R\tau_0\otimes\sigma_1+J_I\tau_2\otimes\sigma_2\    \Rightarrow\  U\H U^\dagger=\left( -J_R\sigma_1+J_I\sigma_2 \right)\oplus\left(-J_R\sigma_1-J_I\sigma_2\right),\\
			\H&=-J_R\tau_0\otimes\sigma_1-J_I\tau_2\otimes\sigma_2\    \Rightarrow\  U\H U^\dagger=\left( -J_R\sigma_1-J_I\sigma_2 \right)\oplus\left(-J_R\sigma_1+J_I\sigma_2\right).
		\end{split}
\end{equation}
From the above mapping of the Hamiltonians, the first blocks of each $ U\H U^\dagger $ represent the systems of two sites with the hopping amplitudes as $ Je^{-i\phi} $, $ Je^{i\phi} $, $ Je^{i(\pi-\phi)} $, and $ Je^{i(\pi+\phi)} $, respectively. Here, we have taken $ J_R=J\cos\phi $ and $ J_I=J\sin\phi $. 
	
\begin{figure}[t]
		\centering
		\includegraphics[scale=0.8]{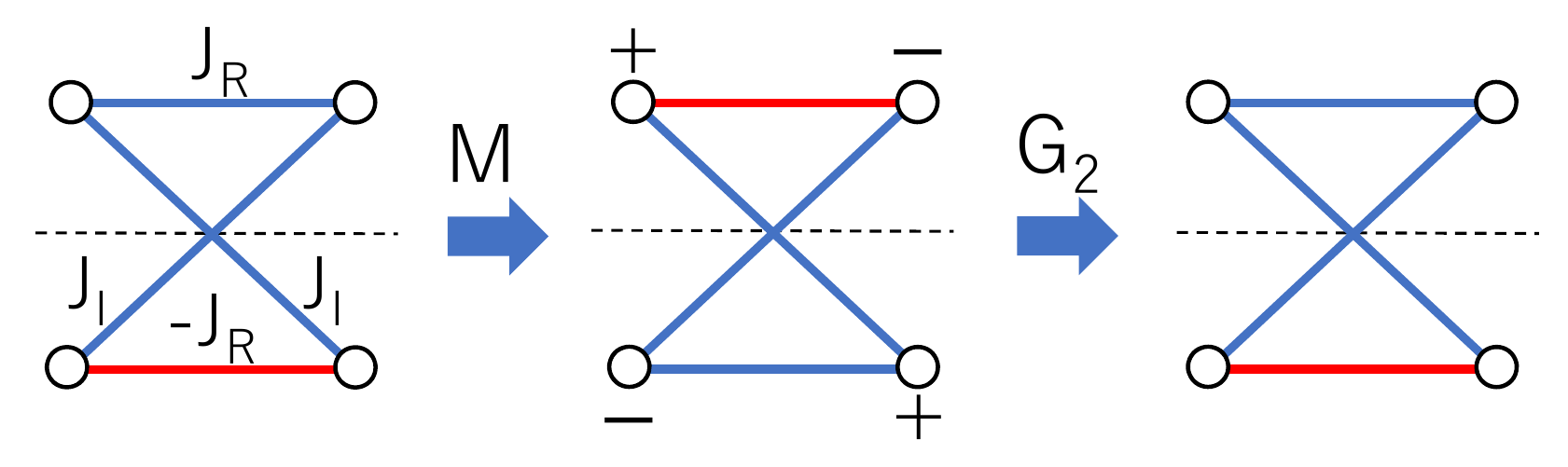}
		\caption{The operation of $ \mathcal{M}_2=\mathsf{G}_2M $.~\label{SM-M2}}
\end{figure}
	
The four twisted $ \pi $-flux blocks in Fig.~\ref{SM-Mapping}(c) are invariant under $ \mathcal{M}_2=\mathsf{G}_2M $ with
\begin{equation}%\label{key}
		\mathsf{G}_2=\tau_3\otimes\sigma_3,\qquad M=\tau_1\otimes\sigma_0.
\end{equation}
Then, we have
\begin{equation}\label{m2}
		\mathcal{M}_2=i\tau_2\otimes\sigma_3,
\end{equation}
by which the algebra $ [\mathcal{T},\mathcal{M}_2]=0,\ \mathcal{M}_2^2=-1 $ of Eq.~(1) in the main text is also satisfied. The operation of mirror symmetry $ \mathcal{M}_2 $ is illustrated in Fig.~\ref{SM-M2}, which takes the second block of Fig.~\ref{SM-Mapping}(c) as an example. Here, time reversal symmetry is represented as $ \mathcal{T}=\mathcal{K} $ with $ \mathcal{K} $ the complex conjugation. For the second block of Fig.~\ref{SM-Mapping}(c), the Hamiltonian can be written as
	\begin{equation}%\label{key}
		\H=J_R\tau_3\otimes\sigma_1+J_I\tau_1\otimes\sigma_1.
	\end{equation}
	Obviously, it satisfies $ [\mathcal{M}_2,\H]=0 $. By the unitary transformation $ U=e^{-i\pi\tau_1\otimes\sigma_3/4 }  $, $ \mathcal{M}_2 $ can be diagonalized as $ U\mathcal{M}_2U^\dagger=i\tau_3\otimes\sigma_0 $. Similarly, the Hamiltonian is transformed into diagonal block as
	\begin{equation}%\label{key}
		U\H U^\dagger=\left(J_R\sigma_1+J_I\sigma_2\right)\oplus\left(-J_R\sigma_1+J_I\sigma_2\right).
	\end{equation}
	Note that time reversal symmetry is now represented in the eigenspace of $ \mathcal{M} $ as
	\begin{equation}%\label{key}
		U\mathcal{T}U^\dagger=-i\tau_1\otimes\sigma_3\mathcal{K}.
	\end{equation}
	Then, for the four twisted $ \pi $-flux blocks in Fig.~\ref{SM-Mapping}(c), the Hamiltonians are transformed as
	\begin{equation}%\label{key}
		\begin{split}
			\H&=-J_R\tau_3\otimes\sigma_1+J_I\tau_1\otimes\sigma_1\ \ \  \Rightarrow\   U\H U^\dagger=\left(-J_R\sigma_1+J_I\sigma_2\right)\oplus\left(J_R\sigma_1+J_I\sigma_2\right),\\	\H&=J_R\tau_3\otimes\sigma_1+J_I\tau_1\otimes\sigma_1\ \ \ \ \  \Rightarrow\  U\H U^\dagger=\left(J_R\sigma_1+J_I\sigma_2\right)\oplus\left(-J_R\sigma_1+J_I\sigma_2\right),\\
			\H&=J_R\tau_3\otimes\sigma_1-J_I\tau_1\otimes\sigma_1\ \ \ \ \  \Rightarrow\  U\H U^\dagger=\left(J_R\sigma_1-J_I\sigma_2\right)\oplus\left(-J_R\sigma_1-J_I\sigma_2\right),\\
			\H&=-J_R\tau_3\otimes\sigma_1-J_I\tau_1\otimes\sigma_1\ \ \  \Rightarrow\  U\H U^\dagger=\left(-J_R\sigma_1-J_I\sigma_2\right)\oplus\left(J_R\sigma_1-J_I\sigma_2\right).
		\end{split}
	\end{equation}
	The first blocks of above resulting Hamiltonians $ U\H U^\dagger $ are the system of two sites with the hopping amplitudes as $ Je^{i(\pi-\phi)}$, $ Je^{i\phi}$, $Je^{-i\phi}$, and $Je^{i(\pi+\phi)} $, respectively.

	\section{The Chern-insulator and mirror-Chern-insulator Triangle Models}
	In this section, we provide detailed information for the triangle-lattice Chern insulator and the corresponding mirror Chern insulator.
	%\subsection{Single-Layer Model}
	
\begin{figure}
		\centering
		\includegraphics[scale=0.8]{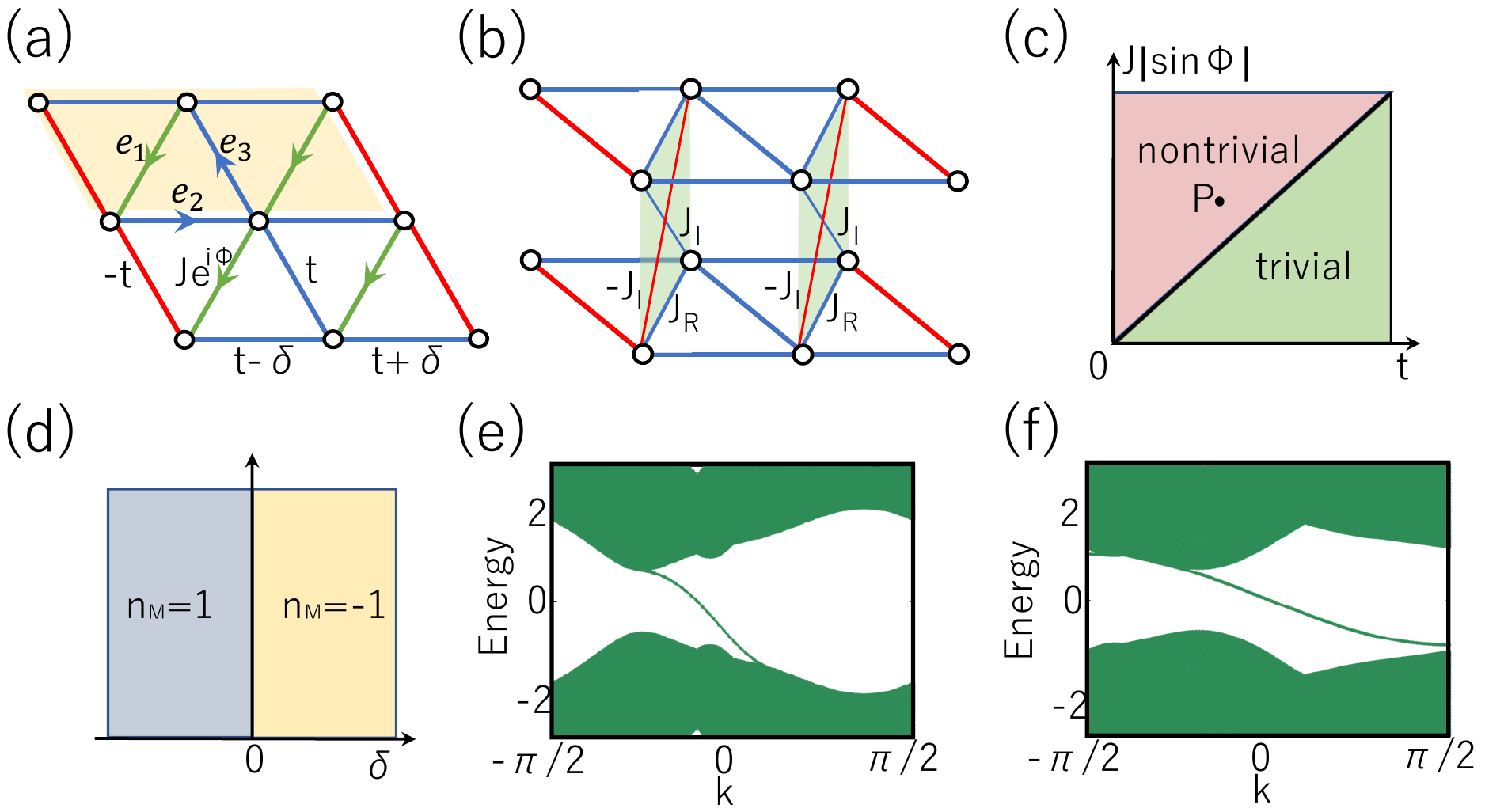}
		\caption{(a) The triangle-lattice Chern-insulator model. (b) The corresponding bilayer mirror-Chern-insulator model. (c) The phase diagram of monolayer system when $ \delta\neq0 $. (d) The phase diagram of triangle-lattice Chern-insulator model when $ t<J|\sin\phi| $. (c) and (d) are the energy spectra for the triangle-lattice Chern-insulator model with edges opened along $ \e_1 $ and $ \e_2 $, respectively. The parameters in calculating (c) and (d) are set as $ t=1 $, $ \delta=0.5 $, $ \phi=2\pi/5 $, and $ J=2 $.~\label{SM-Triangle}}
	\end{figure}

	The monolayer triangle-lattice model is illustrated in Fig.~\ref{SM-Triangle}(a). The three edges of each triangle are $ \e_i $ with $ i=1,2,3 $. The hopping coefficient along each $ \e_1 $ is $ Je^{i\phi} $. The dimerization along $ \e_2 $ is introduced with hopping amplitudes as $ t-\delta $ and $ t+\delta $. Hence, the unit cell consists of two sites as shown by the shadow area in Fig.~\ref{SM-Triangle}(a), and the unit vectors can now be chosen as $ 2\e_2,\ \e_3 $. The hopping amplitude along $ \e_3 $ is $ \pm t $, where the signs $ \pm $ are marked by blue and red lines, respectively. Hence, the Hamiltonian of the monolayer triangular lattice is given as
	\begin{equation}
		h_C(\k,\phi)=[2J\cos(\k\cdot\e_1-\phi)+2t\cos\k\cdot\e_2]\sigma_1+2\delta\sin\k\cdot\e_2\sigma_2-2t\cos\k\cdot\e_3\sigma_3.
	\end{equation}
	The gap of energy spectrum is closed if
	\begin{equation}%\label{key}
		t=J|\sin\phi|,\ \text{or}\ \delta=0.
	\end{equation}
	For $ \phi=0,\pi $ and $ \delta\neq0 $, the monolayer system has time-reversal symmetry (TRS) and the Chern number is zero. When varying $ \phi $, the Chern number remains zero until the gap closes at $ t=J|\sin\phi| $. Thus, the Chern number is nonzero as $ t<J|\sin\phi| $. The phase diagram is shown in Fig.~\ref{SM-Triangle}(c). When $ t<J|\sin\phi| $, the gap closing condition $ \delta=0 $ separates two nontrivial phases with different Chern numbers as indicated in Fig.~\ref{SM-Triangle}(d). The energy spectra in Fig.~\ref{SM-Triangle}(e) and (f) are calculated for the open boundary conditions along $ \e_1 $ and $ \e_2 $, respectively. They demonstrate the topologically nontrivial phase with chiral edge state.
	
	By our general construction of the bilayer model in the main text, the hopping amplitudes of inserted twisted $ \pi $-phase blocks are related to the complex hopping amplitude $ Je^{i\phi} $ as
	\begin{equation}\label{relation}
		J_R+iJ_I=Je^{i\phi}.
	\end{equation}
	Then, the Hamiltonian of this bilayer system can be directly obtained from Eqs.~(9) and (10) of the main text as
	\begin{equation}
		\H^{\text{tr}}(\k)=\begin{bmatrix}
			h_+^{\text{tr}}(\k)  & -ih_-^{\text{tr}}(\k) \\ ih_-^{\text{tr}}(\k) & h_+^{\text{tr}}(\k)
		\end{bmatrix},
	\end{equation}
	where
	\begin{equation}
		\begin{split}
			h_+^{\text{tr}}(\k)&=(2J_R\cos\k\cdot\e_1+2t\cos\k\cdot\e_2)\sigma_1+2\delta\sin\k\cdot\e_2\sigma_2-2t\cos\k\cdot\e_3\sigma_3,\\
			h_-^{\text{tr}}(\k)&=2J_I\sin\k\cdot\e_1\sigma_1.
		\end{split}
	\end{equation}
	From the relation in Eq.~\eqref{relation} and the phase diagram of monolayer system in Figs.~\ref{SM-Triangle}(c) and \ref{SM-Triangle}(d), it is now clear that the phases of bilayer system is irrelevant to $ J_R $ in Fig.~\ref{SM-Triangle}(b). When, $ t<|J_I| $ and $ \delta\neq0 $, the mirror Chern number is nonzero, and the model is a mirror Chern insulator.

	\section{The Hofstadter and mirror Hofstadter models}
	In this section, we provide detailed information for the Hofstadter model and the corresponding mirror Hofstadter model.
	
\begin{figure}[t]
		\centering
		\includegraphics[scale=0.7]{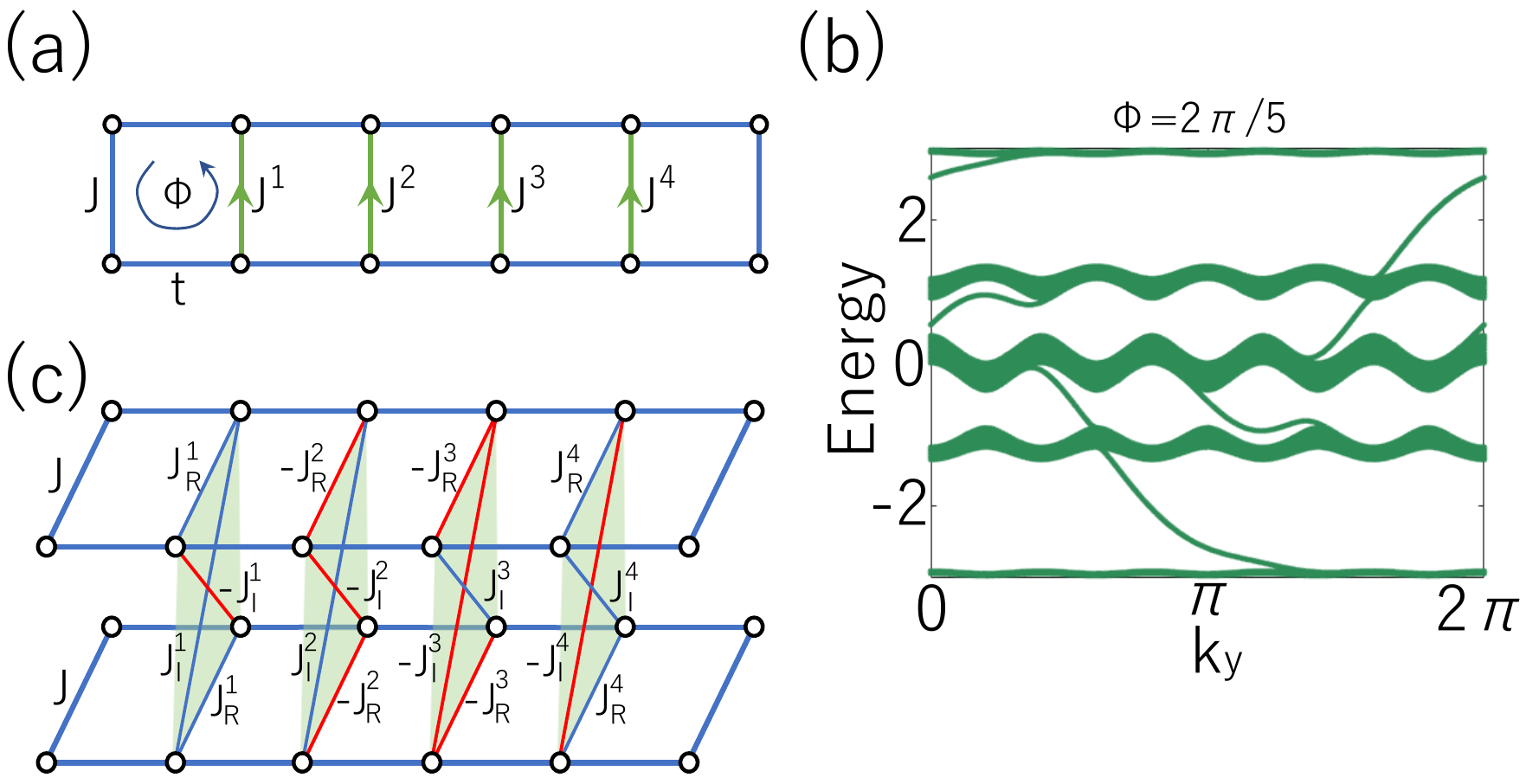}
		\caption{(a) The Hofstadter model with flux $\Phi=2\pi/5$ per plaquette.  The blue edges correspond to positive real hopping amplitudes. The directed green edges denote complex hopping amplitudes: $J^n=Je^{i\frac{2n\pi }{5}}$ with $n=1,2,3,4$. (b) The spectrum of the Hofstadter model with $ t=J=1$, and $\phi=2\pi/5 $. (c) The mirror Hofstadter model. All hopping amplitudes are real, and positive and negative ones are marked in blue and red, respectively.~\label{SM-Square}}
	\end{figure}

	The Hofstadter model with gauge flux $ \phi=2\pi/5 $ through each plaquitte is illustrated in Fig.~\ref{SM-Square}(a). The horizontal hopping amplitude is $ t $, and the vertical one is $ J $. Then, the Hamiltonian is written as
	\begin{equation}%\label{key}
		h_C^{\text{Hof}}=\begin{bmatrix}
			2J\cos k_y & t & 0 & 0 & te^{-ik_x}\\
			t & 2J\cos (k_y-\phi) & t & 0 & 0\\
			0 & t & 2J\cos (k_y-2\phi) & t & 0\\
			0 & 0 & t & 2J\cos(k_y-3\phi) & t \\
			t^{ik_x} & 0 & 0 & t & 2J\cos(k_y-4\phi)
		\end{bmatrix}.
	\end{equation}
	For this system, the Chern numbers of five bands are obtained as
	\begin{equation}%\label{key}
		-1,-1,4,-1,-1,
	\end{equation}
	ordered from higher to lower energies, respectively. As shown in Fig.~\ref{SM-Square}(b), there are four energy gaps with the numbers of chiral edge states as $ 1,2,2,1 $ inside them, respectively. 
	
	By our general method, the Hamiltonian for the mirror Hofstadter model can be written as
	\begin{equation}%\label{key}
		\H^{\text{Hof}}(\k)=\tau_0\otimes h_+^{\text{Hof}}(\k)+\tau_2\otimes h_-^{\text{Hof}}(\k)
	\end{equation}
	with
	\begin{equation*}%\label{key}
		\begin{split}
			h_+^{\text{Hof}}(\k)&=\begin{bmatrix}
				2J\cos k_y & t & 0 & 0 & te^{-ik_x}\\
				t & 2J_R^1\cos k_y & t & 0 & 0\\
				0 & t & -2J_R^2\cos k_y & t & 0\\
				0 & 0 & t & -2J_R^3\cos k_y & t \\
				t^{ik_x} & 0 & 0 & t & 2J_R^4\cos k_y
			\end{bmatrix},\\
			h_-^{\text{Hof}}(\k)&=\begin{bmatrix}
				0 & 0 & 0 & 0 & 0\\
				0 & 2J_I^1\sin k_y & 0 & 0 & 0\\
				0 & 0 & 2J_I^2\sin k_y & 0 & 0\\
				0 & 0 & 0 & -2J_I^3\sin k_y & 0 \\
				0 & 0 & 0 & 0 & -2J_I^4\sin k_y
			\end{bmatrix},
		\end{split}
	\end{equation*}
	where the hopping amplitudes $ (J_R^n,J_I^n) $ with $ n=1,2,3,4 $ of twisted $ \pi $-flux blocks are given in Tab.~\ref{hopping-J}. The bilayer mirror Hofstadter model is illustrated in Fig.~\ref{SM-Square}(c), where the four twisted $\pi$-flux blocks in Fig.\ref{SM-Mapping}(b) are inserted in each unit cell.

	\begin{table}[t]
		\begin{tabular}{|c|c|c|c|}
			\hline  $ (J_R^1,J_I^1) $ & $ (J_R^2,J_I^2) $ & $ (J_R^3,J_I^3) $ & $ (J_R^4,J_I^4) $\\ \hline
			$ (J\cos\frac{2\pi}{5},J\sin\frac{2\pi}{5}) $ & $ (J\cos\frac{\pi}{5},J\sin\frac{\pi}{5}) $ & $ (J\cos\frac{\pi}{5},J\sin\frac{\pi}{5}) $  & $  (J\cos\frac{2\pi}{5},J\sin\frac{2\pi}{5})  $  \\
			\hline
		\end{tabular}
		\caption{The hopping amplitudes of inserted twisted $ \pi $-flux blocks in Fig.~\ref{SM-Square}(c). ~\label{hopping-J} }
	\end{table}

	\begin{figure}[t]
		\centering
		\ \ \ \includegraphics[scale=0.5]{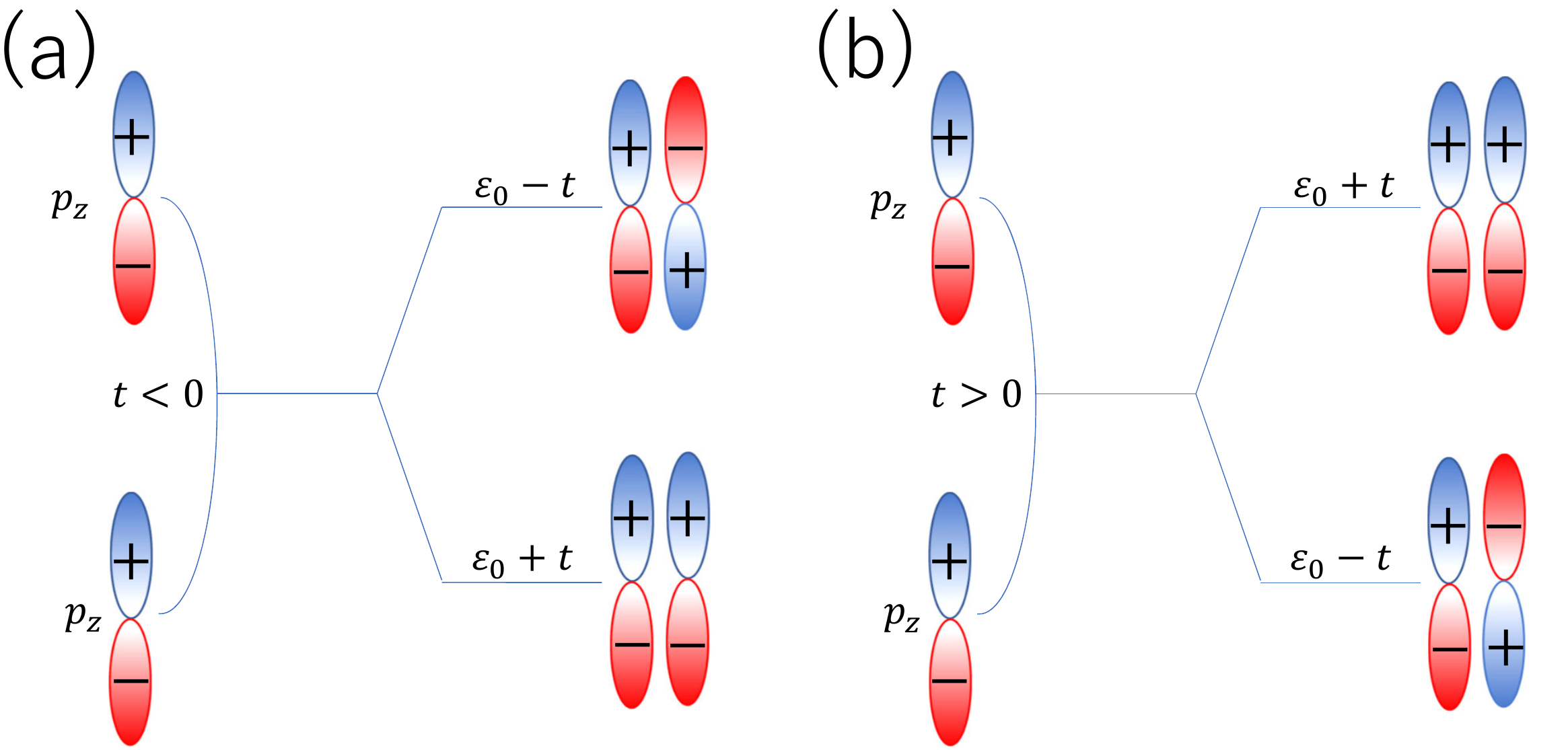}\\   
		\includegraphics[scale=0.5]{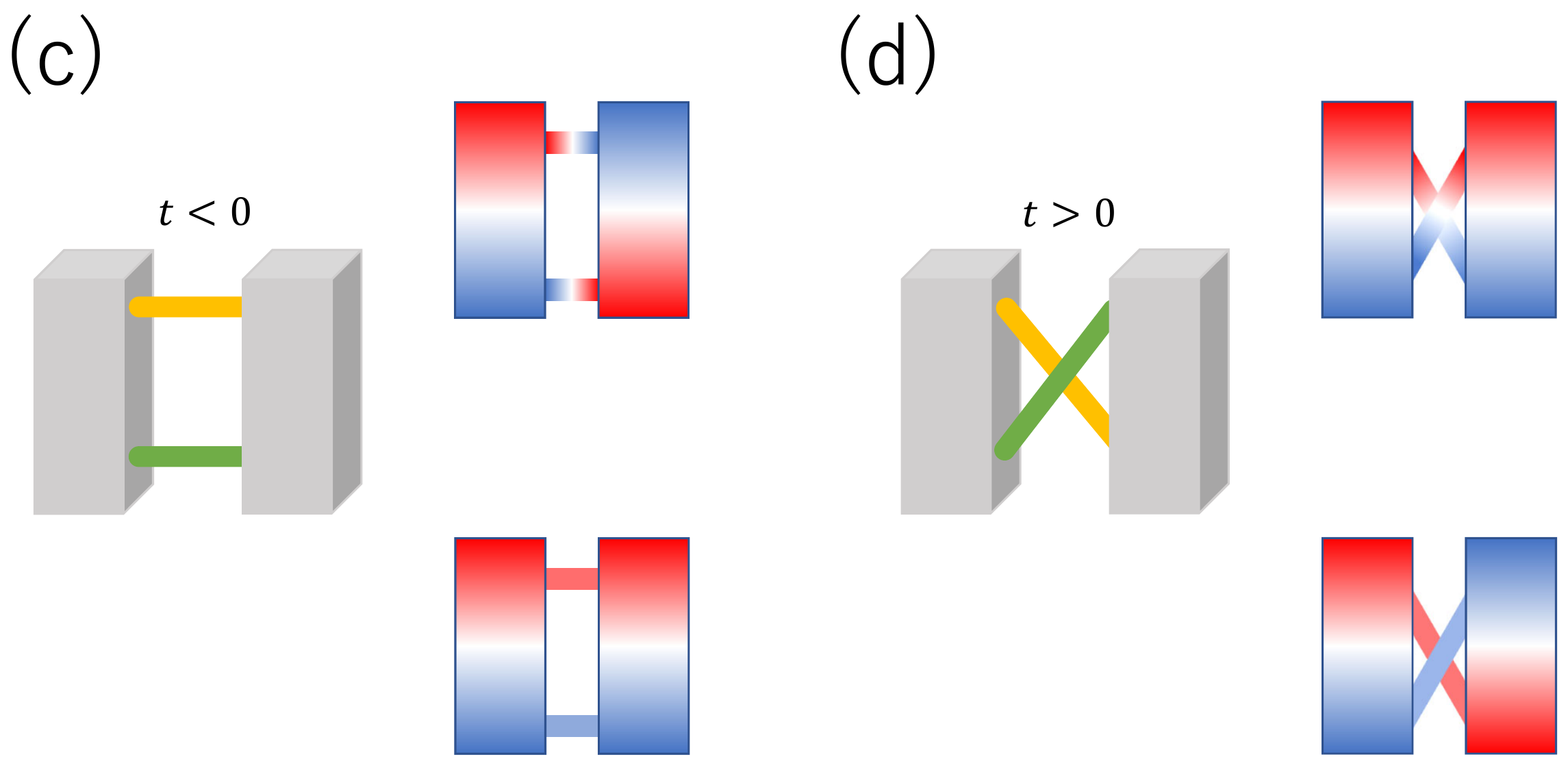}
		\caption{ (a) and (b) illustrate $ \pi $-bonds formed by two $ p_z $ orbitals for $ t<0 $ and $ t>0 $, respectively. For $ t<0 $, the bonding state has lower energy, while the anti-bonding state has higher energy illustrated by the two right configurations of (a). The case for (b) is opposite. (c) and (d) realize the effective coupling $ t<0 $ and $ t>0 $ in acoustic systems, respectively. The gray cubes denote the acoustic resonators connected by the wave guides (green and yellow sticks), and the positive and negative air pressures are marked with blue and red colors.~\label{SM-S5}}
	\end{figure}
	
	\section{The Simulation with Acoustic System}
	
	We review the realization of $ \Z_2 $ gauge field in acoustic systems~\cite{Qi2020,Xue2020,Xue2021}, followed by the proposal for simulating our models as an application. The advantage of using artificial systems rests on the controllable hopping or coupling terms. For example, if there is no coupling between two fixed sites, just remove all the connecting materials between these two sites. For traditional electronic systems, this could be achieved by properly designing the local states with some sorts of symmetries such that the energy integral vanishes. To simulate electronic system by acoustic system, the eigen oscillation mimics the electronic wave function in solid systems. In this way, the equation of motion for electrons is simulated by the dynamic equation of the oscillation in acoustic system with the frequency playing the role of energy. To be precise, lattice sites in solid systems are replaced by acoustic resonators, while hopping between different sites is realized by coupling between different resonators. By connecting the acoustic resonators with wave guides or coupling tubes, the coupling amplitude can be easily controlled since it is completely determined by the radius of wave guide or coupling tube.
	
	\begin{figure}[t]
		\centering
		\includegraphics[scale=0.5]{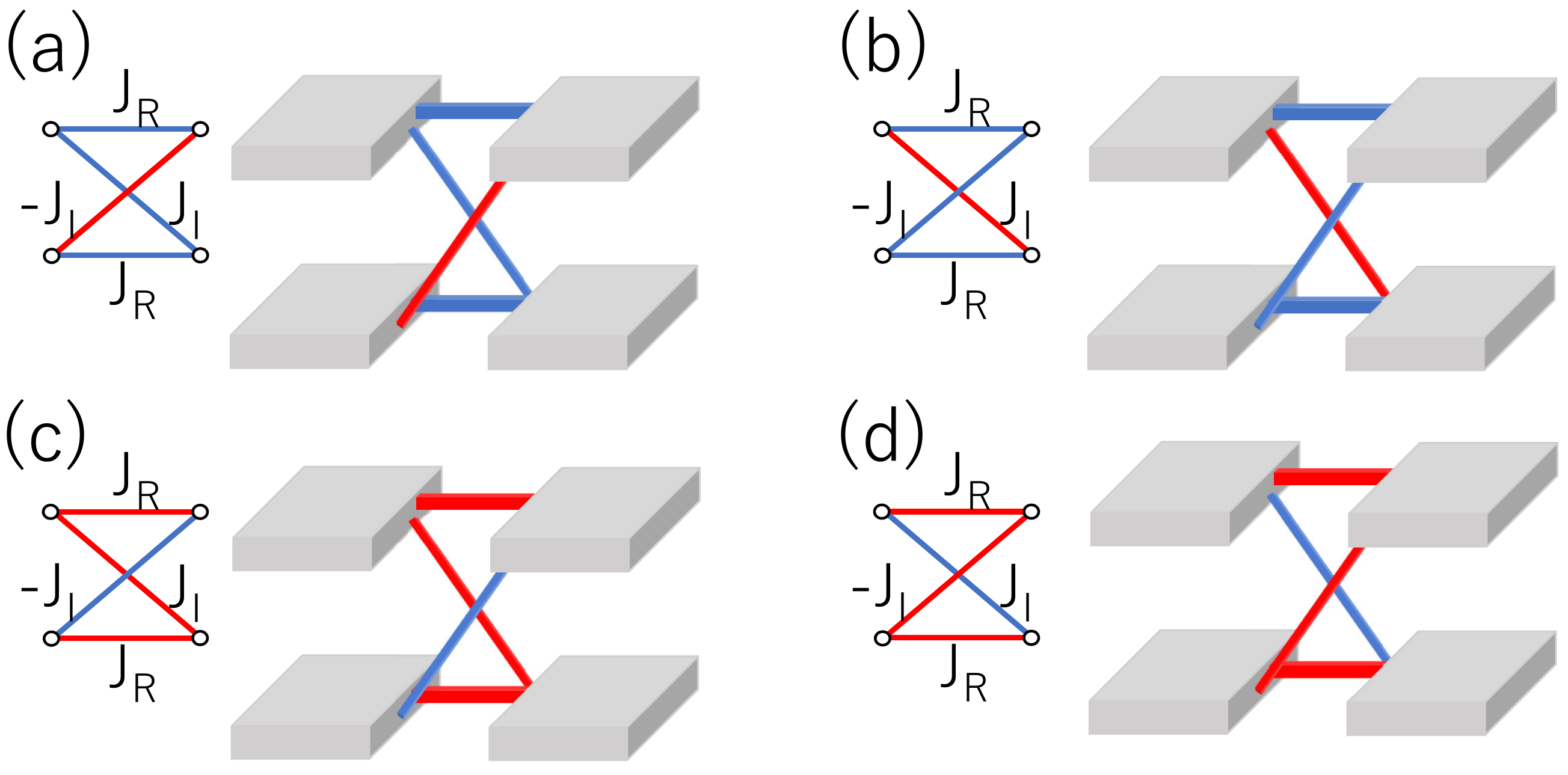}
		\caption{ Realization of building blocks in Fig.~\ref{SM-Mapping}(b). The red and blue links connecting different acoustic resonators represent the connection structures in Fig.~\ref{SM-S5}(c) and \ref{SM-S5}(d), respectively. ~\label{SM-S6}}
	\end{figure}
	
	To control the sign of the coupling between different resonators, the dipolar mode far away from other modes of the resonator is chosen, which resembles the atomic $ p $-orbital as shown in Fig.~\ref{SM-S5}. So, let's begin with the coupling between two remote $ p_z $ orbitals in solid system as shown in Fig.~\ref{SM-S5}(a) and \ref{SM-S5}(b). Assuming the coupling coefficient is $ t $ and the onsite energy is $ \varepsilon_0 $, the Hamiltonian of this small system can be written as
	\begin{equation}\label{H-SM}
		H=\begin{bmatrix}
			\varepsilon_0 & t \\ t & \varepsilon_0
		\end{bmatrix}
	\end{equation}
	with two eigenstates
	\begin{equation}%\label{key}
		\ket{\pm}=\frac{1}{\sqrt{2}}\begin{bmatrix}
			1 \\ \pm1
		\end{bmatrix}
	\end{equation}
	for $ E_{\pm}=\varepsilon_0\pm t $, respectively. If $ t<0 $, the low-energy state is bonding state while the high-energy one is anti-bonding state as shown by Fig.~\ref{SM-S5}(a). The case for $ t>0 $ is just opposite as shown in Fig.~\ref{SM-S5}(b). As to the acoustic resonator, the dipolar mode manifests the sinusoidal distribution of air pressure inside the resonator, which just mimics the wave function of $ p_z $ orbital. For hopping simulated by coupling different resonators with wave guides, there is a physical picture here. The wave guides introduce the perturbation to eigen modes of the resonators. In acoustic systems, the uniform distribution of air pressure always has lower frequency or lower energy than the sinusoidal distribution. Therefore, by considering the perturbation from the wave guides in Fig.~\ref{SM-S5}(c), the bonding state has lower frequency than the anti-bonding one. From the lesson of Fig.~\ref{SM-S5}(a) that the coupling coefficient is negative if the bonding state has lower energy, we obtain the effective coupling $ t<0 $. Namely, there is a $ \pi $ hopping phase for the connection structure of Fig.~\ref{SM-S5}(c). The case of Fig.~\ref{SM-S5}(d) is just opposite to that of Fig.~\ref{SM-S5}(c). In this way, the $ \Z_2 $ gauge field can be readily realized in acoustic systems by properly designing the connection structures.

	The above strategy of controlling the coupling between acoustic resonators, as a proven technology, has been widely used in topological acoustic systems. We now turn to the realization of our models. It is clear that once the building blocks in Fig.~\ref{SM-Mapping} are simulated, our models are immediately realized. We illustrate the realization of the building blocks in Fig.~\ref{SM-Mapping}(b) by Fig.~\ref{SM-S6}, and the realization of building blocks of Fig.~\ref{SM-Mapping}(c) is just similar. In Fig.~\ref{SM-S6}, the red and blue connections between different acoustic resonators represent the connection structures of Fig.~\ref{SM-S5}(c) and \ref{SM-S5}(d), respectively. As to the realistic experimental setup, the mature 3D-printing technique with photosensitive resin has been demonstrated to be a powerful method of realizing all these models in acoustic systems.

	%the eigen modes of each isolated resonator mimic the atomic orbits of each site in solid system. To control the sign of coupling between these resonators, the dipolar modes with the frequency $ \omega_0 $ are considered here, which are like the $ p $-orbits for electrons.

	%the states of the dipolar modes in acoustic systems represent these of the electronic wave function in solid system. Hence, the equation of motion for electrons in solid systems is simulated by the dynamic equation of the oscillation in acoustic system. The lattice sites of electronic system are now replaced by the acoustic resonators with properly designed resonating frequencies as the onsite energies. The hopping between different sites is now simulated by placing the wave guide or coupling tube between them. 

	%\subsection{Electric Circuit}

	%For the simulation by electric circuits, the voltages of circuit nodes represent the wavefunctions of local states for electronic systems, and the dynamics of circuits just plays the role of the Schr\"{o}dinger equation for electrons. In this way, the spectrum of oscillation modes inside the circuit mimics the energy spectrum of electronic system.

%\bibliographystylesupp{apsrev}
%\bibliographysupp{SM-mirror-symmetry}

\end{document}